\documentclass[runningheads]{llncs}
 
\usepackage{algorithm}
\usepackage{algpseudocode}
\usepackage{booktabs}
\usepackage[T1]{fontenc}

\usepackage{graphicx}

\begin{document}
\title{A Fair and Lightweight Consensus Algorithm for IoT}

\author{Sokratis Vavilis\orcidID{0000-0002-3104-3973} \and
Harris Niavis\orcidID{0000-0002-5941-9987} \and
Konstantinos Loupos\orcidID{0000-0002-9162-7233}}
\authorrunning{S. Vavilis et al.}

\institute{Inlecom Innovation, Athens, Greece \\
\email{\{sokratis.vavilis,harris.niavis, \\ konstantinos.loupos\}@inlecomsystems.com}}
\maketitle              
\begin{abstract}

With the rapid growth of hyperconnected devices and decentralized data architectures, safeguarding Internet of Things (IoT) transactions is becoming increasingly challenging. Blockchain presents a promising solution, yet its effectiveness depends on the underlying consensus algorithm. Conventional mechanisms, such as Proof of Work and Proof of Stake, are often impractical for resource-constrained IoT environments. To address these limitations, this work introduces a fair and lightweight hybrid consensus algorithm tailored for IoT. The proposed approach minimizes resource demands on the nodes while providing a fair and secure agreement process. Specifically, it utilizes a distributed lottery mechanism to ensure fair block proposals without requiring dedicated hardware. In addition, to enhance trust and establish finality, a reputation-based voting mechanism is incorporated. Finally, we experimentally validated the key features of the proposed consensus algorithm.

\end{abstract}
\section{Introduction}

As we enter an era of hyperconnected devices and decentralized data architectures, maintaining the integrity and security of transactions within such a distributed ecosystem presents significant challenges \cite{SICARI2015146,8462745}. 
Factors such as device heterogeneity, insecure communication protocols, and limited computational resources create barriers to establishing trust in the IoT device lifecycle \cite{yousefnezhad2020security}.
Integrating blockchain into IoT systems offers a transformative approach to enhance security, data integrity, and interoperability \cite{fi12090157,7467408}. 
Especially in the last-mile delivery sector, where blockchain can enhance trust between urban logistics stakeholders through green smart contracts and digital identities \cite{urbaine}, enabling transparent and tamper-evident coordination across the supply chain. At the core of this integration is the consensus algorithm, a fundamental component that dictates network throughput, transaction validity, and immutability. Developing efficient and resilient consensus mechanisms is essential to enable secure, scalable, and trustworthy decentralized IoT ecosystems.

Although fundamental consensus algorithms such as Proof of Work (PoW) and Proof of Stake (PoS) have been effective in general-purpose blockchains such as Bitcoin and Ethereum, they face unique challenges when applied to IoT ecosystems.
The resource-intensive nature of PoW, for example, poses a significant hurdle in the resource-constrained environments characteristic of IoT devices \cite{wen2020blockchain,TRIPATHI2023100344,stefanescu2022systematic}. Moreover, the need for energy-efficient consensus mechanisms becomes paramount as the proliferation of connected devices continues unabated.

This paper introduces a novel, fair and lightweight consensus algorithm tailored
for the demands of Blockchain in IoT scenarios. The algorithm addresses limitations of existing mechanisms by optimizing for low resource consumption and fairness. 
It uses distributed lottery mechanisms to ensure equal participation and reputation-based block voting to foster trust in the process, all without depending on specific hardware, enabling its applicability to various devices.

This approach improves performance, security, and trust in IoT environments, such as decentralized last-mile delivery IoT networks, where secure and efficient sharing of logistics events between stakeholders is critical for traceability, non-repudiation, and regulatory compliance. By addressing key limitations of existing consensus mechanisms, the proposed solution can further promote broader blockchain adoption in real-world IoT and logistics applications.

The remainder of this paper is organized as follows. Section ~\ref{related_work} introduces the reader to blockchain consensus algorithms by discussing related work. In Section 3 we present an overview of the proposed algorithm along with details of each phase, while Section 4 analyzes the experiments performed to assess the algorithm in terms of fairness and robustness to attacks.
Finally, Section 5 concludes and points out
directions for future work.

\section{Related work} \label{related_work}

Consensus algorithms are regarded as the cornerstone of blockchain as they provide the means for nodes to agree on the state of the blockchain. Specifically, they provide the mechanisms by which the nodes agree on the blocks to be added to the blockchain. The characteristics of the consensus algorithm are crucial for both the performance and the security properties of the blockchain solution. Various approaches have been proposed in the literature, originating from the field of distributed network systems and, more recently, from blockchain technology per se. 
Generally, consensus algorithms can be classified into two broad categories: voting-based and proof-based \cite{JAIN2025100065}. 

Traditional distributed systems approaches have been adapted for permissioned blockchains, as they share similar foundational assumptions. These methods rely on a voting mechanism, where nodes elect a leader and vote on block proposals, with final decisions determined by majority rule. Their primary goal is to ensure consistency and finality, guaranteeing that all nodes have a unified view of the blockchain state after each round. Nonetheless, these approaches tend to be more centralized and face scalability issues due to the high communication overhead required for consensus. Additionally, many voting-based models assume that most participating nodes are both honest and consistently available.

Over the past decades, Paxos \cite{lamport1998part} was the dominant voting-based consensus mechanism, though its inherent complexity hindered its adoption. To address this issue, Raft \cite{ongaro2014search} was proposed as a more understandable alternative.
Raft enhances clarity by introducing a straightforward leader election mechanism, in which each node has the chance to campaign for a Leader role for a given round. This is achieved by random time-based rounds during which a candidate solicits votes from their peers, requiring a majority to become the leader.
Both approaches, although crash-fault tolerant, assumed the honesty (i.e., trust) of all participating nodes, restricting their applicability in many real-world scenarios. To this end, Byzantine fault-tolerant (BFT) solutions were suggested. One of the most widely applied and efficient mechanisms of this kind is the Practical Byzantine Fault Tolerance (pBFT) \cite{castro2002practical}, which was designed to tolerate up to |(n-1)/3| malicious nodes. It uses a three-phase protocol, in which failing to reach consensus among all consensus nodes triggers the election of a new primary node (leader). The major drawback of pBFT is its quadratic complexity $O(n^2)$ compared to the linear complexity of the other two algorithms. Modern and scalable voting-based methods have also been proposed exclusively for blockchain networks. A prominent example is Proof of Vote \cite{8291964}, in which a committee of predefined members elects butler nodes, which are responsible for validating transactions and adding blocks to the chain.

Public permissionless blockchains face distinct security challenges, as they cannot assume trustworthy nodes and must prioritize performance due to their large scale. Their design emphasizes decentralization, security, and availability, often at the cost of immediate consistency and finality, leading to blockchain forks. To mitigate forks and achieve eventual consistency probabilistically, these blockchains employ mechanisms such as computational challenges (e.g., solving cryptographic puzzles) or leveraging node properties (e.g., stake ownership).

Regarding public blockchains, PoW \cite{dwork1993,bitcoin2008bitcoin} is the most popular consensus algorithm, as it is the one adopted by Bitcoin. PoW is a fully decentralized consensus mechanism that requires network members to expend computational effort to solve a mathematical puzzle. The first node to solve the puzzle adds a block to the chain. Verifying a solution to such a puzzle can be easily and undeniably confirmed by other nodes. PoW methods, however, experience low throughput and have a considerable environmental impact due to their high energy consumption. To overcome the latter, different mechanisms have been proposed in the literature, such as Proof of Capacity / Proof of Storage  \cite{ateniese2014proofs,dziembowski2015proofs}, which pose challenges related to data storage, which has a lower environmental footprint. Other approaches focus on physically selecting nodes employing the notion of randomness similar to participating in a lottery. For instance, in Proof of Elapsed Time \cite{poet} each node has a random timer, and the node that times out first adds a block to the chain.  Similarly, in the Proof of Luck (PoL) algorithm \cite{milutinovic2016proof}, each node calculates a random number and the luckiest (e.g., smallest number) adds a block. To guarantee the honest behavior of the nodes and the indisputability of the random procedures, a Trusted Execution Environment (TEE) must be present in each node.
Based on the above, it becomes apparent that these solutions require nodes to have a considerable number of resources or features, which renders them inappropriate for an IoT context \cite{wen2020blockchain,TRIPATHI2023100344,stefanescu2022systematic,SAHRAOUI2025}. Nonetheless, recent approaches like Proof of Verifiable Function (PoVF) utilize Verifiable Delay Functions (VDFs) and Verifiable Random Functions (VRFs) to achieve unpredictable node selection without relying on specialized hardware~\cite{xiong2025povf}.

Apart from methods that depend on a physical challenge to select a node to add a block, other consensus algorithms make such a choice based on node properties. For instance, Proof of Authority \cite{poauth} solutions rely on a set of pre-authorized nodes whose identity and trust have already been verified. In each round, one node from that set is chosen to add a block, while the rest of the nodes in the network follow that decision. Although quite efficient, these approaches increase the level of centralization, and authorized nodes usually must be pre-approved in the physical world \cite{wen2020blockchain,TRIPATHI2023100344}. 
The most prominent consensus algorithm in the aforementioned family is PoS \cite{king2012ppcoin}, which is quickly gaining ground among modern cryptocurrencies. In this approach, the node to add a block is selected considering the stake (e.g., coins held) that a node has in the network. The underlying intuition is straightforward: the higher the stake, the greater the chances of being chosen to add a block. In this context, high-stakes nodes have an incentive for the network to operate as designed and be profitable. Moreover, nodes that do not fulfill their responsibilities may lose a portion of their stake (i.e., slashing) as punishment; thus, they are less likely to misbehave. PoS algorithms exist in many different flavors, such as Delegated PoS \cite{dpos}, which aim to be more efficient, or Pure PoS \cite{gilad2017algorand,algorand2} solutions which aim to be more secure and fair. The main drawback of such approaches is that they are designed for cryptocurrencies where nodes have an economic incentive, making their application difficult in different contexts \cite{wen2020blockchain,TRIPATHI2023100344}. Lastly, such algorithms favor nodes with significant investments in the network, leading to increased centralization.

Although narrow in their applicability, PoS algorithms have been proven to be secure, efficient, and have limited resource requirements, thus gaining the interest of the research community. To overcome their applicability problems, researchers have worked on expanding their concept to cover a wider variety of properties. 
To this end, Proof of Importance (PoI) \cite{niavis2022consenseiot} algorithms have been proposed. This family of algorithms extends the concept of stake by considering other aspects that highlight the importance of a node in the network. For example, the contribution to the network, the number of successful transactions, and the trustworthiness of the nodes are considered.  A particular type of PoI algorithm is Proof of Trust / Reputation algorithms \cite{zou2018proof,kang2019toward,sun2021rc}  that assess the overall trustworthiness of nodes (e.g., based on their characteristics and past behavior) to select a node for adding a block to the chain. These methods show great potential, and there is substantial ongoing research in the field of IoT. However, the main challenge with these algorithms is related to their fairness, as nodes with higher importance tend to dominate the blockchain network.

In real-world applications, most consensus algorithms adopt a hybrid approach, combining different mechanisms to capitalize on their strengths while mitigating their shortcomings. One such example is Proof of Activity \cite{bentov2014proof}, which integrates elements of both PoW and PoS. In this model, a lightweight PoW process is used to mine a block, after which PoS is used for validation and block addition.
Other notable hybrid approaches include Ethereum’s PoS and Algorand \cite{gilad2017algorand,algorand2}, which, while primarily PoS-based, incorporate pBFT to enhance security and guarantee finality. Hybrid consensus mechanisms have also been explored in the IoT domain, such as PoEWAL \cite{andola2020poewal}, which combines PoW with PoL. Given their ability to balance security, efficiency, and scalability, hybrid approaches present a promising direction for IoT applications.

While we have already discussed the applicability of fundamental consensus paradigms in IoT, several recent studies proposed algorithms exclusively for IoT \cite{SAHRAOUI2025}. For example, 
\cite{biswas2019pobt} extends the HyperLedger Fabric framework  
with a novel approach to reduce the nodes participating 
in the consensus process, enhancing its suitability for IoT environments. 
In \cite{dorri2021tree}, a leader-based consensus algorithm is introduced, where leaders precompute a hash-based token (i.e., consensus code) and are assigned to manage transactions associated with it. 
This approach allows validators to create and commit blocks containing transactions under their supervision, thereby 
reducing the need for extensive inter-node coordination.

Lastly, two works share similarities with our algorithm: REVO \cite{barke2024revo} and LVCA \cite{verma2025lvca}. REVO leverages reputation to form trusted committees that elect nodes for block addition, whereas LVCA employs a consensus based on random lottery voting among trusted nodes to select the block proposer. In contrast, our algorithm emphasizes fairness by randomly selecting block proposers from the global pool of nodes while using reputation only to validate and add blocks. 

\section{Fair and lightweight consensus}

In this section, we discuss our consensus algorithm for IoT devices. First, we provide an overview of our proposal, and then we proceed to discuss the details of each phase of the algorithm.

\subsection{Overview}
Blockchain offers strong potential to address IoT security challenges due to its decentralized and immutable nature. However, the application of consensus algorithms in IoT remains a challenge. Resource-heavy mechanisms such as PoW are unsuitable for constrained devices, whereas lighter alternatives like PoS or PoI may unfairly favor nodes, effectively demotivating newcomers. Voting-based methods also struggle, as they often incur high communication overhead or lack resilience against malicious actors, especially in dynamic, public networks.

\begin{figure}[ht] 
  \centering
  \includegraphics[width=0.83 \linewidth, scale=0.60]{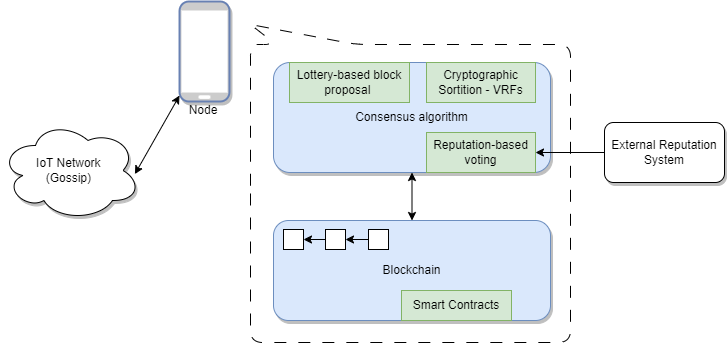}
  \caption{Consensus algorithm structure overview}
  \label{building_blocks}
\end{figure}

To address these limitations, we present a novel hybrid consensus algorithm for the IoT that is both lightweight and fair. As illustrated in Figure~\ref{building_blocks}, the algorithm comprises two main parts: a distributed lottery mechanism for block proposals and a reputation-based voting process to finalize the agreement on a specific block.  The lottery mechanism employs Verifiable Random Functions (VRFs), enabling nodes to generate a random lot in a verifiable and decentralized fashion. The voting process relies on a consortium of trusted nodes who vote for the best block proposal observed in a given round. The trust level of each node is determined through an external reputation system ~\cite{vavilis2014reference,fortino2020trust,aaqib2023iot}. We note that our approach is not constrained to a particular reputation model; additionally, for the purposes of this work, we assume the existence of such a system.

Unlike other consensus algorithms, our approach suits resource-constrained environments by avoiding heavy computations and special hardware requirements like Trusted Execution Environments (TEE) as done, for instance, in PoL \cite{milutinovic2016proof}. It ensures fairness, giving all nodes equal opportunity to propose blocks regardless of their resources. A trusted consortium of randomly selected high-reputation nodes vets only the best proposal per round without influencing block creation itself. Their incentive to preserve reputation for future committee and network participation discourages malicious behavior.

\subsection{Algorithm flow}

\begin{figure}[h]
  \centering
  \includegraphics[scale =0.28]{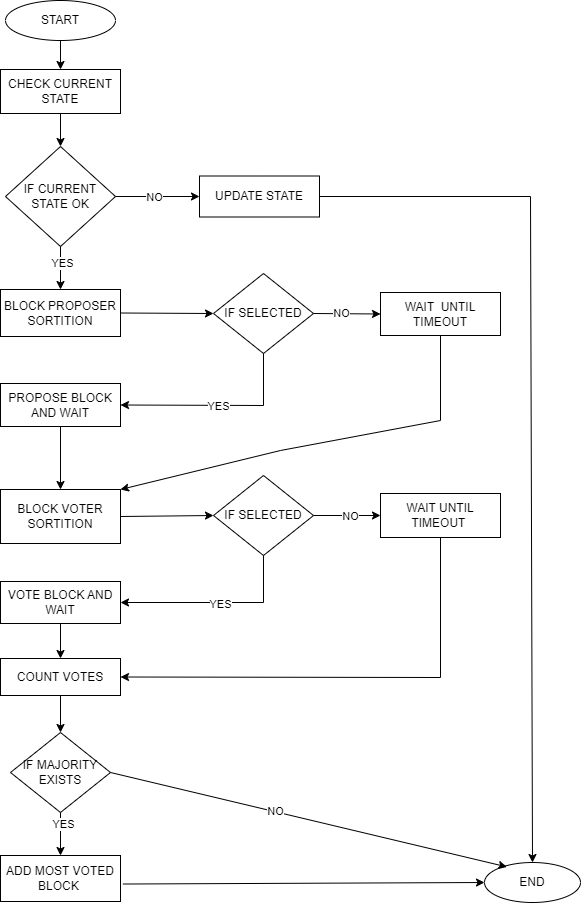}
  \caption{Consensus algorithm execution flow}
  \label{flow}
\end{figure}

The operational flow of the proposed algorithm is shown in Figure \ref{flow}. In particular, the nodes of the network compete with each other in adding a specific block to the chain by continuously executing the depicted process. At the beginning of each round (i.e., order of block to be added), nodes check whether they are in the correct round. This is achieved by assessing whether the node has received any messages regarding a round greater than the current height of the local blockchain known to the node. In case the node is out of sync with the other nodes, it requests from its neighboring nodes the most up-to-date chain available. Based on replies from its neighbors, the node adopts the longest valid chain (in terms of blocks).  Otherwise, the nodes proceed with the block proposal. 

During the block proposal phase, each node executes a cryptographic sortition algorithm to determine whether it is selected to propose a block. If chosen, the node generates a proposal using the randomly computed lot created in the sortition process and broadcasts it. The node then waits for a predefined timeout period to allow message dissemination across the network. In the subsequent voting phase, the cryptographic sortition mechanism is applied again to select high-reputable nodes for the voting committee, which votes for the best block proposal (i.e., the one with the highest lot value) observed. After the voting timeout expires, each node determines whether a majority vote has been achieved for a proposed block, which is then added to the blockchain. Notably, if the number of participating nodes in each phase is known (e.g., in permissioned networks or through the reputation system), nodes can proceed to the next phase as soon as a majority is reached, without waiting for the full timeout. A reduced version of the algorithm, focused on permissioned networks but omitting sortition and dynamic timeouts, was previously implemented and validated in the Hyperledger Fabric platform~\cite{michalopoulos2024integrating}.

During this process, each node executes in parallel a message handling routine, implementing a gossip protocol responsible for disseminating messages in the network. For efficiency reasons, previously seen messages and messages regarding rounds smaller than the currently known by a node are discarded. To protect the authenticity and integrity of the messages exchanged, they are cryptographically signed. We note that the communication approach used is quite common in consensus algorithms and peer-to-peer networks, and its further investigation is out of the scope of this work.

\subsection{Algorithm Details}
In the following, we describe the proposed consensus algorithm in detail.

\paragraph{\textbf{Random Lotteries and Sortition}}

Random lotteries are central to our design, governing both block proposal and voting. Thus, we require a secure and verifiable method for generating randomness in a distributed setting. While a plethora of solutions exist \cite{choi2023sok,raikwar2022sok,randao,micali1999verifiable}, few suit resource-constrained IoT environments.

In our algorithm, we have chosen VRFs \cite{micali1999verifiable} to generate random lots, as it is a mature primitive with low computational and communication overhead. VRFs rely on public key cryptography to produce verifiable and uniform random numbers, in the form of hashes, given an input. In particular, a node can use a common public value as input to the VRF, known as the seed, and using its secret key will generate a random number, along with a cryptographic proof. 
Other nodes in the network can verify the validity of the produced lot using the accompanying proof, the seed used in its generation, and the public key of the lot issuer.
Each round seed is calculated as $seed_r = VRF_{sk}(seed_(r-1)||role)$, where $r$ represents a round. To properly utilize the properties of VRFs, the publicly known seeds used in each round must be chosen at random and not controlled by potential threat agents. 
To this end, we require that all agents' secret keys are defined well in advance the seed of a particular round (e.g., in the previous round or earlier). In this respect, we follow a similar strategy as the one discussed in \cite{gilad2017algorand}, and we recommend the reader to read that work for more details.

\begin{algorithm}
\caption{Cryptographic Sortition}\label{sortition}
\begin{algorithmic}[1]
\Procedure{Sortition}{$sk,role,seed$}
\State $lot,proof \gets VRF_{sk}(seed||role)$
\If{$role$ is $PROPOSAL$ \textbf{and} $lot \ge (THRESHOLD_{proposal} \times MAX\_LOT)$}
    \State \textbf{return} $\textbf{True}, lot, proof$
  
\ElsIf{$role$ is $VOTE$  \textbf{and} $node\_reputation \ge THRESHOLD_{reputation}$ \textbf{and} $lot \ge  (THRESHOLD_{vote} \times MAX\_LOT$)}
\State \textbf{return} $\textbf{True}, lot, proof$
\Else
    \State \textbf{return} $\textbf{False}$
\EndIf

\EndProcedure
\end{algorithmic}
\end{algorithm}

As the number of participants in the network can be arbitrarily high, we need a mechanism to select random subsets of nodes to actively participate in the block proposal and voting process. This will effectively decrease the communication overhead introduced by exchanging a large number of block proposals or voting messages. To meet this requirement, we employ cryptographic sortition based on VRFs, to secure the random selection of nodes. Due to the use of VRFs, cryptographic sortition is reliable and verifiable by the nodes of the network.  As shown in Algorithm \ref{sortition}, of a particular round and phase of the algorithm (i.e., block proposal or voting), a node calculates a random lot using the VRF and if this value is above a predefined threshold, then the node is selected for participation.  Other nodes of the system can verify the outcomes of the sortition mechanism for a node using Algorithm \ref{sortition_ver}.  In the following, we discuss how the sortition mechanism is used for proposing blocks and voting. 

\paragraph{\textbf{Block Proposal}}

In each round, every node executes the sortition algorithm (Alg. \ref{sortition}) to assess whether it is chosen to propose a block. Initially, the node calculates a random lot for the round using the VRF function along with its public key and the round's seed. If the calculated lot is over a specific threshold, the node proceeds to send its block, including the generated lot with the associated proof and the block's seed. The nodes receiving the block proposal employ Algorithm \ref{sortition_ver} to verify it. This algorithm verifies the VRF and then uses similar criteria to the ones used in the sortition. We note that the proposed sortition algorithms are designed to give equal chances of selection to every legible node, without prioritizing nodes based on any other factor 
than the random lot.

The threshold aims to limit the number of nodes proposing a block, hence limiting the number of proposals in the network. To calculate the threshold value, we use a weight (i.e., $THRESHOLD_{proposal}$) and multiply it by the maximum potential value of the VRF (i.e., $MAX\_LOT$). Since VRF's output is uniformly distributed, the $THRESHOLD_{proposal}$ defines the anticipated ratio of nodes to be selected. For instance, by setting $THRESHOLD_{proposal} = 0.9$, we anticipate that only 10\% of the nodes will be chosen. 

We note that using this approach, the number of selected block proposers grows linearly to the number of nodes, although at a lower rate defined by $THRESHOLD_{proposal}$. However, in case the total number of nodes in the network is known, it is possible to dynamically adjust the $THRESHOLD_{proposal}$ value to fix the number of block proposers per round. 
Albeit trivial in a permissioned blockchain network, it is not the case for public permissionless networks. Nevertheless, assessing the potential number of nodes in a permissionless network is beyond the scope of the current work. 

Finally, to limit the amount of information transmitted over the network, during the block proposal phase, only the generated lots along with their proofs and the proposed block's hash are sent over the network. Only after the end of the proposal phase is the actual winning block(s) transmitted. Next, we argue that due to the random selection procedure of our approach, potential adversaries have limited chances to be selected, which is relative to their ratio over the participants of the network. In practice, an adversary has to control over 50\% of the network to have meaningful chances of being chosen and posing a threat in the operation of the blockchain. Such a scenario would be feasible in a permission-less network via Sybil-attacks since our algorithm lacks explicit protection against Sybil attacks.  However, even in that case, attackers will only be able to slow down the rate of adding blocks to the chain. Such an effect could be mitigated by using the reputation system to detect patterns of malicious behavior and ban nodes that consistently misbehave. We leave the investigation of such an approach for future work. Lastly, we note that the safety of the blockchain would not be compromised due to the reputation-based voting mechanism.

\begin{algorithm}
\caption{Cryptographic Sortition Verification}\label{sortition_ver}
\begin{algorithmic}[1]
\Procedure{VerifySortition}{$pk,role,seed,lot,proof$}
\If{$VerifyVRF_{pk}(lot,proof,seed||role)$ is $\textbf{False}$}
\State \textbf{return} $\textbf{False}$
\EndIf
\If{$role$ is $PROPOSAL$ \textbf{and} $lot \ge (THRESHOLD_{proposal} \times MAX\_LOT)$}
    \State \textbf{return} $\textbf{True}$
  
\ElsIf{$role$ is $VOTE$  \textbf{and} $node\_reputation \ge THRESHOLD_{reputation}$ \textbf{and} $lot \ge (THRESHOLD_{vote} \times MAX\_LOT)$}
\State \textbf{return} $\textbf{True}$
\Else
    \State \textbf{return} $\textbf{False}$
\EndIf
\EndProcedure
\end{algorithmic}
\end{algorithm}

\paragraph{\textbf{Voting Phase}}

After the block proposal phase, selected nodes of the network vote for the best block proposal, defined as the one with the highest lot value seen in the current round.
To this end, every node of the network executes the sortition algorithm (Alg. \ref{sortition}) to decide whether they are part of the voting committee. The internal mechanics of the sortition algorithm for the voting phase are quite similar to the block proposal phase. In particular, VRFs are again used for the calculation of a random lot, however, there are a few notable differences.

First and foremost, we have introduced a reputation-based selection criterion. The intuition behind this choice is to only select nodes with a high enough reputation, that have already proven their good behavior in the network. To achieve this, a $THRESHOLD_{reputation}$ value must be set to a value that semantically denotes a high reputation. For instance, if a reputation system employs a $[0,1]$ scale, the threshold value could be set to $0.8$. We stress the fact that $THRESHOLD_{reputation}$  should be high enough, in order not to allow adversaries to be selected for voting. We also argue that with a high enough $THRESHOLD_{reputation}$ value, potential attackers would have to invest a significant amount of time and effort to achieve and retain a high enough reputation. Moreover, such a strategy should be done for multiple nodes, to achieve a high enough probability to be chosen by the sortition algorithm and form a majority. 

Secondly, we introduce a different threshold ($THRESHOLD_{vote}$) for the lots used in the voter selection process. The reason for such a choice is that we want to adjust the size of voting committees in a different manner than for the block proposal. In addition, since the number of highly reputable nodes is known, this threshold value could be set to result in a voting committee of fixed size, further limiting the chances of potential adversaries being selected.

\paragraph{\textbf{Block addition}}

When the voting phase is completed, the network nodes will decide which block they will add to the blockchain. In particular, nodes first verify the received voting messages using the Algorithm \ref{sortition_ver}. Then if a block has gathered the majority of the votes, it is added to the chain. Otherwise, the nodes do not add any blocks and proceed to the next round. It should be noted that nodes will only add blocks containing valid data (e.g., valid transactions). 
Assuming that the votes originate from high-reputable and trustworthy nodes, this approach ensures finality and safety of the chain. It should also be noted that due to the use of digitally signed messages and verifiable lots, the integrity of the exchanged information is not at stake. Therefore, potential adversaries may only be able to hinder the efficiency of the network by reducing the rate at which blocks are added to the chain.

\section{Experiments and Discussion}

This section presents the experiments conducted to evaluate the proposed consensus algorithm. We begin by outlining the objectives of the tests and the experimental setup, followed by a detailed discussion of the results obtained.

\subsection{Experiment objectives and setting}

Our primary objective is to evaluate the performance of the proposed consensus algorithm concerning its fairness property and its robustness to attackers. To this end, we have developed a reference implementation of the algorithm in Python. To enable communication and message exchange between the nodes, we implemented a REST API using Flask. Over this API a simple gossip protocol was developed, allowing nodes to spread messages to five random neighbors. Avoidance of repeating already sent messages, such as block proposals, was implemented to reduce communication overhead. It is important to note that in terms of performance, Python and Flash are sub-optimal choices. However, their ease of use enables rapid development which fits our objective of creating a Proof of Concept (PoC) to assess the algorithm’s fundamental properties. 

Regarding the configuration of the consensus algorithm, we made the following choices. Similarly to other algorithms \cite{milutinovic2016proof,gilad2017algorand}, we set the overall round timeout period to 20 seconds (12 seconds for the block proposal and 8 seconds for voting). Empirically, this time window would be sufficient for spreading information (i.e., messages) over the network.
Furthermore, to reduce the network overhead, we also set $THRESHOLD_{proposal} = 0.9$, allowing only 10\% of the nodes to propose a block in each round. Regarding the voting mechanism, we set $THRESHOLD_{reputation} = 0.8$ (in the range [0,1]) to allow only high-reputable nodes to vote, while $Threshold_{vote}$ is set to allow a voting committee of 11 nodes. The nodes obtain reputation information via a reputation system. We also note that we set the size of the blocks at 200 kilobytes. Lastly, to reduce the computation overhead introduced by cryptographic operations (e.g., digital signatures of messages), we replaced such operations with less computationally intensive alternatives (e.g., time delays).

For the experiment, we deployed our sample implementation in the AWS cloud infrastructure, deploying up to 100 nodes (depending on the experiment use-case) in a similar number of t3.micro instances. Of these nodes, 15 are highly reputable, with a reputation of 0.9, from which the voting committee members will be randomly selected. Note that despite that the number of highly reputable nodes does not have an impact on the experiment results, it should be high enough to allow the formation of a voting committee. We assume that these nodes remain honest throughout the experiments.  We also acknowledge that our experiment setting and implementation are not optimized for computational and network efficiency. However, we argue that this has a limited impact on our evaluation since we study the behavior of our algorithm with respect to fairness and robustness to adversaries rather than for efficiency and throughput.

\subsection{Assessing the Fairness property}

A key aspect of our design is the fairness with which nodes add blocks to the chain. To evaluate this aspect, we measure the diversity of the block proposers (i.e., the number of different proposers) that have successfully added a block to the chain. We also measure the size of the globally accepted chain (i.e., the longest most common chain) and the number of nodes that are in sync with it. 

We evaluate the fairness property with a set of experiments deploying different numbers of nodes. We let the consensus algorithm run for ten minutes for each number of nodes. We repeat this process five times and report the averages per metric. The results of our experiments are shown in Table \ref{tab:fair}.

\begin{table}[h]
\centering
\caption{Fairness experiments results}
\label{tab:fair}
\addtolength{\tabcolsep}{2pt}    
\begin{tabular}{@{}lllll@{}}
\toprule
Nodes           & 25 & 50 & 75 & 100 \\ \midrule
Blocks added    & 30   & 30   & 30   &  30   \\
Proposer diversity & 23   & 26  &  26  &  27  \\
Nodes in sync   &  25  & 50   &  73  &  94 \\ \bottomrule
\end{tabular}
\addtolength{\tabcolsep}{1pt}    
\end{table}

The first finding of our experiment is that the algorithm maintains a stable throughput of three blocks per minute, regardless of the number of nodes.
Despite the relatively small-scale setup (up to 100 nodes) compared to large public blockchain networks with thousands of participants, the results indicate that the proposed approach performs as expected, with a round-time of approximately 20 seconds.
Furthermore, most nodes stay synchronized with the global chain, though a slight drop in synchronization is observed as the network size increases. After inspecting the node logs, we found that the desynchronized nodes were lagging for a block, which could be the result of network discrepancies. These nodes are expected to catch up with the last block during the next round. Lastly, regarding the fairness property of our consensus algorithm, we notice that the block proposer diversity grows from 76\% to 90\% as the number of nodes increases. This behavior is expected because the uniform distribution of the probability of the lotteries is better expressed on a greater scale.

\subsection{Robustness to attacks}

Before we move on to discuss how we assess the robustness of attacks, we need to define the attacker's profile. In this regard, we consider that an attacker can deploy multiple malicious nodes that will not respect the consensus protocol.
Since attackers cannot influence the integrity of the exchanged messages and cannot forge the random lots (due to the use of VRFs), they concentrate their efforts on selecting which information to disseminate over the network to affect the block added to the chain. Malicious nodes may selectively forward block proposals and votes aiming to stop the expansion of the chain with new blocks. Next to that, malicious nodes aim to increase the chances that a malicious node is chosen to propose a block, effectively halting the addition of a block for that round. In our experiments, we assume that malicious nodes cannot have a high reputation and thus cannot participate in the voting committee. 

To assess the robustness of our algorithm against attackers, we employ metrics similar to those used in the previous section. In particular, we measure how the presence of attackers impacts the fairness and expansion of the chain. We do this with a set of experiments deploying 100 nodes and a varying ratio of malicious nodes on the network. For each case, we let the consensus algorithm run for ten minutes. We note that by definition, the malicious nodes are deliberately not in sync with the chain.  The results of our experiments are shown in Table \ref{tab:attack}

\begin{table}[h]
\centering
\caption{Robustness to attacks results}
\label{tab:attack}
\addtolength{\tabcolsep}{2pt}  
\begin{tabular}{@{}llllllll@{}}
\toprule
Malicious Nodes & 0   & 10  & 20  & 30  & 40  & 50 \\ \midrule
Blocks added    & 30  & 26  &  26  & 25   &  18  &   15     \\
Proposer diversity & 27  & 23   & 23   & 22  &  17   &   14    \\
Nodes in sync   & 94  & 84  & 75  & 68  &  58  &   47     \\ \bottomrule
\end{tabular}
\addtolength{\tabcolsep}{1pt}  
\end{table}

Evaluating the results, we observe that the ratio of malicious nodes has an impact on the throughput of the blockchain network. In particular, we notice a small drop of 14\% in the number of blocks added when the ratio of attackers is relatively low (10\% to 30\%), however, a more significant drop of 40\% -50\% is noticed when the number of malicious nodes represents almost half of the nodes. This behavior can be justified by the fact that the ratio of malicious nodes in the network is proportional to the probability of being chosen to propose a block. Such an effect could be mitigated by using the reputation system to decrease the reputation of nodes that consistently misbehave, and eventually ban them from the network. We leave the investigation of this use case for future work.  We note that regardless of the drop in overall throughput, the network manages to operate with both safety and finality. Last but not least, it is noticed that the block proposer diversity (i.e., fairness) and the (honest) nodes synchronized to the global chain are not affected by the presence of adversaries in the network. 
This is evident in the scenario with 50 malicious nodes, where 14 of 15 blocks were proposed by different nodes, and 47 out of 50 nodes remained synchronized.

\section{Conclusion \& Future Work}

This work introduced a fair and lightweight hybrid consensus algorithm designed for IoT environments, balancing security, fairness, and efficiency. It combines random lotteries for fairness with reputation-based voting for safety and finality, all with minimal resource demands. 
 A sample implementation validated the algorithm's fairness property and its resilience against adversarial attacks. However, real-world efficiency and scalability in IoT remain open for further investigation. Finally, future research could further leverage the reputation mechanism to improve security and performance without compromising fairness.

\section*{Acknowledgment}

This work has received funding from the EU Horizon Europe Programme in the framework of the "Upscaling Innovative Green Urban Logistics Solutions Through MultiActor Collaboration and PI-Inspired Last Mile Deliveries" project (URBANE), under GA Number 101069782

%
%
%

 \bibliographystyle{splncs04}
\bibliography{sample-base}
\end{document}